\documentclass[letterpaper,twocolumn,10pt]{article}

\usepackage{enumitem}
\setlist{nolistsep}
\usepackage{todo}
\usepackage{float}
\usepackage{booktabs,dcolumn,caption}
\usepackage{xcolor}
\usepackage{colortbl}
\usepackage{usenix,epsfig,endnotes}
\usepackage{hyperref}

\colorlet{tablerowcolor}{gray!10} 
\newcommand{\rowcol}{\rowcolor{tablerowcolor}} %
\newcommand{\code}[1]{\textsf{#1}}

\begin{document}

%


\date{}







\title{\Large \bf Platform-Centric Android Monitoring---Modular and Efficient}

\author{
{\rm Jan-Christoph K\"uster}\\
NICTA and Australian National University
\and
{\rm Andreas Bauer}\\
TU Munich, Germany, and NICTA
} 


\maketitle

\thispagestyle{empty}

\subsection*{Abstract}

We present an add-on for the Android platform, capable of intercepting
nearly all interactions between apps or apps with the platform,
including arguments of method invocations in a human-readable format.  A
preliminary performance evaluation shows that the performance penalty of
our solution is roughly comparable with similar tools in that area.  The
advantage of our solution, however, is that it is truly modular in the
sense that we do not actually modify the Android platform itself, and
can include it even with an already running system.  Possible uses of
such an add-on are manifold; we discuss one from the area of runtime
verification that aims at improving system security.



\section{Introduction}
\label{sec:intro}
In a nutshell, monitoring a system means gaining state information while
it executes, for example, for the purpose of profiling (cf.\
\cite{Yoon2012}) or security hardening (cf.\
\cite{EGC+2010,XSA2012,BGH+2013}), the latter being also our motivation
although our technical contribution is entirely application-agnostic.
For Android, efficient monitoring relies on having access to low-level
(and under normal circumstances inaccessible) system events, such as
Linux system calls, for example, that are generated by the running apps.
%
%
Existing monitoring approaches are roughly distinguishable as being
either \emph{app-} or \emph{platform-centric}: while the former
basically rewrite and repackage apps under scrutiny in order to
intercept the relevant events and app interactions,
the latter are usually known to rely on modifications to some key system
components
of the Android platform, 
such as its API, virtual machine (Dalvik VM), the various system
services, C/C++ libraries, the OS kernel, and whatever else is needed
for an application at hand. 
In this paper, we present a 
monitoring add-on for Android, which is platform-centric, yet does
\emph{not} require such comprehensive system modifications.

As app-centric monitoring does not even require root access
to the Android platform, it, 
arguably, 
has an edge over the platform-centric approaches described in the
literature, as far as usability goes.
For example, AppGuard \cite{BGH+2013}, which manages the modification
and repackaging process of untrusted apps entirely on a user's ``off the
shelf'' phone
has been downloaded more than one million times, hinting at the fact
that it is not just used by a select group of domain experts---on the
contrary  \cite{BGH+2013}.
However, the ease in usability comes at the expense of some inherent
vulnerabilities, namely that the added security controls and event
interception code are actually being executed within the very same
(Java) app they are meant to keep an eye on.
AppGuard redirects method calls in the Java layer by altering method
references in the Dalvik VM, but cannot take note of system calls that
are happening deeper down the stack.
%
Aurasium \cite{XSA2012}, another app-centric monitoring tool, works by
redirecting low-level function pointers away from an app's dynamically
linked Bionic C library to a security monitor, which only gives users
the desired security benefits if the modified app does not provide its
own C library---for whatever reason.
%
%
Besides, 
decompiling and changing 
third-party apps usually violates the licence agreement and 
destroys their original signatures and therefore the ability to automatically update in the future.
However, AppGuard solves this particular problem by also taking control
over updates for the ``patched'' apps.
%







The platform-centric approaches described in the literature (cf.\
\cite{EGC+2010, DBLP:journals/corr/GunadiT13, Bugiel2012, Hornyack2011,
  NKZ2010}) 
generally tie deeper into the platform, and so they need to find ways to
cope with 
the extensive market fragmentation surrounding the Android platform
(cf.\ \cite{HZF+2012}); that is, they have to make sure that their
changes are ported to various versions of the OS as well as to different
hardware platforms, some of which do not lend themselves well to running
custom-built Android versions, due to closed-source device drivers, for
example.
%
%
%
%
TaintDroid \cite{EGC+2010} is a pioneering platform-centric tool for
taint flow analysis, which requires modifications beginning from the OS
kernel all the way up to the Dalvik VM.  Although it is being actively
ported to newer versions of Android since its inception, users of
vendor-specific Android releases may find it difficult to use, unless
they are sufficiently experienced to not only compile their own version
of Android, including the TaintDroid changes, but also to make it work
on a hardware platform of their choice.
On the other hand, these deep ties into the platform enable analyses
that go beyond what is possible with app-centric tools, since
information can be tracked down to the OS kernel level as is needed for
taint flow analysis.

\textbf{Our approach}, conceptually, is a combination of the advantages
of app- and platform-centric monitoring; that is, to offer a software
add-on that can be loaded even into a currently running Android system,
yet is able to trace app interactions all the way down to the OS kernel
level.
%
Technically, this functionality is mostly provided by an OS kernel
module, which uses Linux' own kernel debugging facility (called
\code{kprobes}\footnote{https://www.kernel.org/doc/Documentation/kprobes.txt})
to intercept system calls (see also \S\ref{sec:droidTracer}).  As such,
our approach is strictly platform-centric and does not just apply to
specifically prepared apps, but everything that is executed on the
system.  
In fact, we are able to intercept Android's own inter process
communication (IPC) mechanism called
Binder,\footnote{http://developer.android.com/reference/android/os/Binder.html}
and can therefore gather information on almost all possible interactions
between apps and the Android platform or via the platform.
However, as platform-centric approach, we too require root privileges,
if only to load the module.  This may seem restrictive, but one should
keep in mind that it has become somewhat common practice to obtain root
privileges on unlocked devices in order to use custom ROMs like
CyanogenMod,\footnote{http://www.cyanogenmod.org} install utilities like
firewalls or packet filters, or to be able to edit system files.
Our add-on is available as open source software under the moniker
DroidTracer from \url{http://kuester.multics.org/DroidTracer/}.

In the remainder, we will shed some light on DroidTracer's internals
(\S\ref{sec:droidTracer}), how it may be put to good use
(\S\ref{sec:rv}), and discuss its performance overhead
(\S\ref{sec:exps}).  Some conclusions are given in \S\ref{sec:conc}.
\section{How It All Works}
\label{sec:droidTracer}
In a nutshell, our add-on implements two different but complementary
means of event monitoring: firstly, it directly intercepts relevant
Linux system calls and secondly, it intercepts IPC calls via Android's
internal communication infrastructure called Binder (see below).

\paragraph{System Call Interception}

%
Normally, Android apps, which are executed on a Java virtual machine and
without root privileges, do not directly trigger system calls.  Instead
they request via a Java API that some service be used, or some resource
be allocated.
These high-level requests eventually translate into a sequence of system
calls that ultimately make the kernel open a file, send something to a
(virtual) device, etc.
Our kernel module contains handler methods that get invoked by small
bits of code (so called probes) that are dynamically inserted by
\code{kprobes} at almost arbitrary kernel addresses, corresponding to
the respective system calls.
For example, to check if an app eventually reads from (or writes to) a
file on the SD card, we break into the system call \code{sys\_open()}
and check if its argument, \code{filename}, which always holds the full
path of a file to open, contains the sub-string ``sdcard''.
Similarly, we can track down every internet connection established by an
app, independent of the API version in use, by checking in
\code{sys\_connect()} if the socket address family is of type
\code{AF\_INET} (resp.\ \code{AF\_INET6}), and so forth.

Alternatively, we could have placed hooks directly into the high-level
service handlers or the various system services of the platform, like
\cite{DBLP:journals/corr/GunadiT13, Bugiel2012, Hornyack2011, NKZ2010}
do, in order to be notified when relevant events take place.  Although
this is possible and sometimes unavoidable, as in the case of
TaintDroid, since taint flow analysis requires \emph{seamless} tracking
of control and data flows, this has the disadvantage that such hooks are
specific to particular platform releases and, generally, not portable.
Besides, the functionality provided by our kernel module can be entirely
switched on or off without rebooting or modifying the system.
%
%
%

\paragraph{Binder Interception}
Binder is Android's main communication infrastructure between apps and
system services and it implements the popular remote method invocation
(RMI) scheme. Essentially it lets an object running in one Dalvik VM
invoke methods on an object running in another Dalvik VM.
\begin{figure*}[ht]
  \centering
  \includegraphics[width=0.85\textwidth]{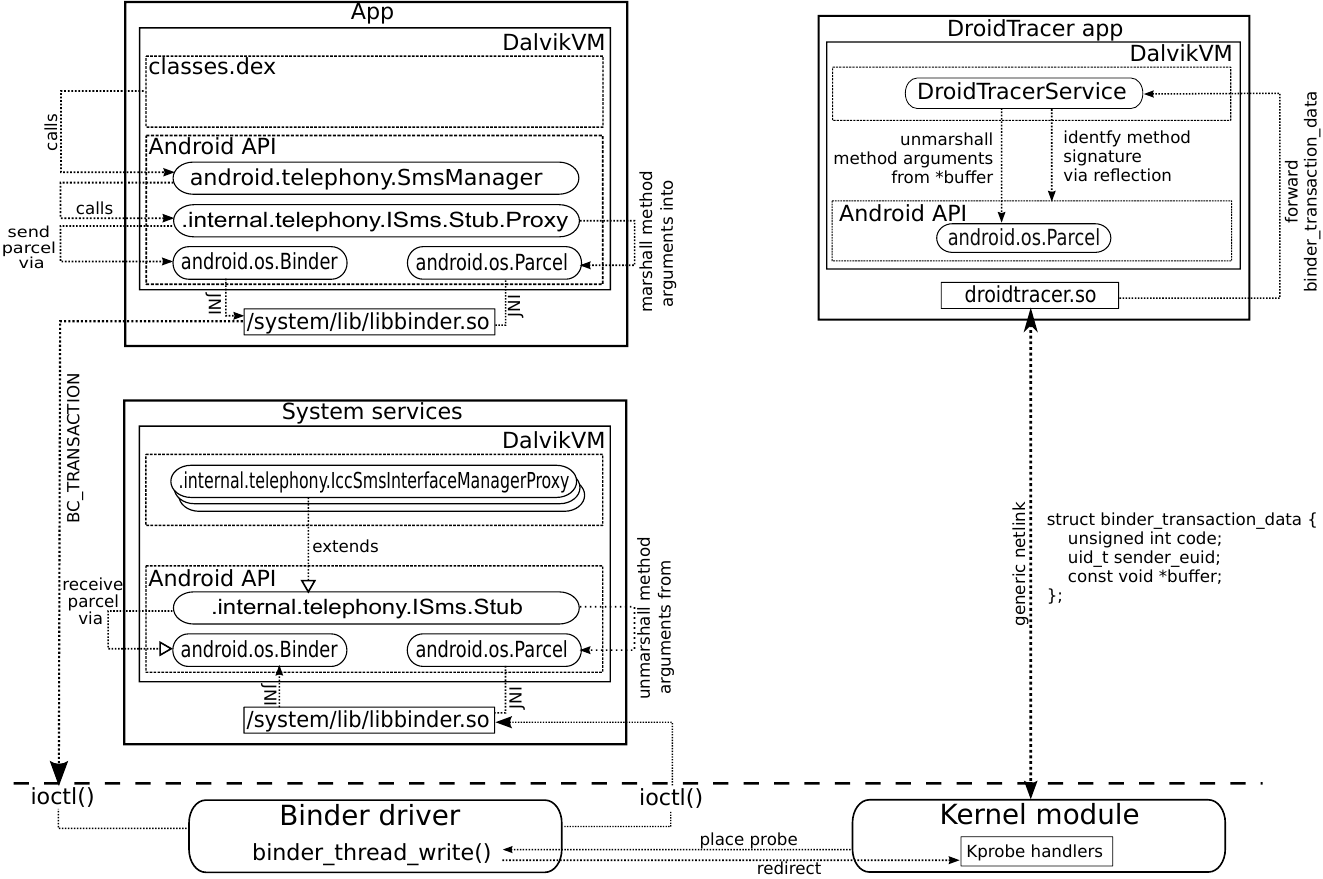}
  \caption{An app accessing Android's system services via Binder (left)
    and our add-on (right).}
  \label{fig:rmi}
  \vspace{-1em}
\end{figure*}
This communication is exemplified in Fig.~\ref{fig:rmi}. It shows an app
that attempts to send an SMS by using methods provided by the class
\code{SmsManager}. This class, however, is more of an interface rather than the
actual implementation of the service which does the sending. What is
actually happening is that the \code{SmsManager} invokes a so called proxy
which compactly encodes arguments and data types of
\code{SmsManager::sendTextMessage()}, for efficient transmission via the
Binder kernel driver. The class \code{android.os.Parcel} is the corresponding
container for this data, also providing the required encoding and
decoding methods. The parcel is then received by
\code{com.android.internal.telephony.ISms.Stub}, which is the counterpart of
the proxy and which takes care of the decoding.  The stub, in turn,
invokes \code{IccSmsInterfaceManagerProxy::sendText()} with the arguments originally
intended for \code{sendTextMessage()} to tell the GSM driver (not depicted) to
do the physical transmission of said message.

The communication with Binder basically follows a stringent
communication protocol between sender and target.  By hooking into
the function \code{binder\_thread\_write()} of its kernel driver, we are
able to tell when a new communication session commences; that is, when a
sender has triggered the transmission of the \code{BC\_TRCANSACTION}
signal.  Specifically, this signal (defined inside \code{binder.h} along
with other protocol signals) indicates that the custom C structure,
\code{binder\_transaction\_data} (also defined inside \code{binder.h}),
has been initialised at a set address in memory.  It contains, amongst
others, information on the sender, method arguments, and the method
name---although compactly and not human-readable encoded (see paragraph
on data extraction below).
We read this information from the structure before the kernel copies it
into the address space of the target process.


Naturally, one can also intercept system events and messages later in
the control flow; for example, when the
\code{IccSmsInterfaceManagerProxy} communicates with the kernel driver
of the radio device.  At this late stage, however, one would lose the
information about the original sender app of an SMS, since its Linux
User ID (UID), which uniquely identifies an app on the Android platform,
is not passed to the kernel.  In other words, we would see that an SMS
is being sent, but not by whom.
As system services check the permission of the client based on its UID,
Android uses Binder to communicate said UID directly to the service
(by writing the field \code{sender\_euid} inside
\code{binder\_transaction\_data}).  Via this indirection, it is not
possible for a rogue process to directly query a service, using a fake
UID, for example.
These observations suggest to us that there are no real alternatives to
intercepting system events and communication at the level of Binder, if
one is interested not only in the occurrence of certain events, but also
in the associated meta data.

\paragraph{Data Extraction}
Our interception of Binder lets us monitor every API method invocation,
but name, types and data of arguments as well as caller and callee are
either dropped or compactly encoded (i.e., marshalled) in terms of
abstract IDs for efficiency reasons.
Therefore, we also need to concern ourselves with unmarshalling the
data once intercepted; that is, determine which process/app tries to
send an SMS, what its content is, who the recipient is, etc.
%

To that end, we have realised \code{DroidTracerService} (see right upper
corner of Fig.~\ref{fig:rmi}), a user space app without its own GUI. It
serves mainly two purposes: On the one hand, it acts as an
interface between the kernel module and Android's ``unmarshalling
algorithms, on the other hand, it is able to pass the intercepted system
events on to a client app (cf.\ \S\ref{sec:rv} for an example).
As such it is similar to a classical system library, which provides an
API for third-party apps to use rather than a dedicated GUI.

Unmarshalling Binder communication, however, is not something that
is typically done by apps. In fact, there is no public API for this task
nor any complete documentation on the marshalling algorithms that
would allow users to devise their own implementations easily.  Hence, we
reverse engineered this whole process to some extent, to be able to use
the official unmarshalling methods of Android's
\code{android.os.Parcel}. This has the advantage that our
solution ends up using the same methods that are being used by the
Binder communication anyways, and is therefore not specific to
particular versions of the Android OS.
%

We discovered that \code{android.os.Parcel} provides a method to fill its internal
data structure from a byte array. Thus, we can create a
\code{Parcel} object directly from the intercepted field \code{buffer}
inside \code{binder\_transaction\_data}, to apply the Android specific
unmarshalling methods to it, namely, \code{readString()},
\code{readInt()}, \code{readFloat()}, etc.
However, we cannot access arguments easily, as a \code{Parcel} object
for efficiency is missing the information about the order in which they
appear; that is, we have to invoke the unmarshalling methods in the
right oder to read the encoded arguments from it correctly. In the
offical unmarshalling process, the stub knows how its proxy has encoded
the arguments using the counterpart methods \code{writeString()},
\code{writeInt()}, \code{writeFloat()}, etc. By examining several proxy
classes, we then observed they usually marshall the method arguments in
the same order they appear in the corresponding method's signature.  We
can access the signature and with it the ordering and the types of the
arguments via reflection, but only under the prerequisite of having the
called interface and method name revealed and at hand first. For our
example in Fig.~\ref{fig:rmi} this information is
\code{com.android.internal.telephony.ISms} and \code{sendText()}.
We found out that the interface name is always encoded first (before the
method arguments) in a \code{Parcel} object; this is, for the stub to
read and ensure that the proxy called its correct counterpart. We can
access it by applying \code{writeString()}. In terms of decoding the
called method name, \code{binder\_transaction\_data} leaves us with an
integer, called \code{code}.
%
A closer look at several stub classes reveals that the mapping of
\code{code} to its according method name is, in fact, defined there.
For example, in \code{com.android.internal.telephony.ISms.Stub} the
method \code{sendText()},
is encoded in terms of a variable
%

{\centering
{\tt static final int TRANSACTION\_sendText = 5}.
}

\noindent
If the according proxy class sends this \code{code} through Binder, the
stub uses it to
first trigger the correct unmarshalling and second, the execution of
\code{sendText()}.
%
As the suffix of \code{TRANSACTION\_sendText} equals the
method name, we are able to use reflection to extract it.
%

Marshalling methods exist only for Java primitives, and some other
types, such as Binder references, primitive arrays, etc. But more
complex objects can be sent through Binder, too; that is, if they
implement the method \code{createFromParcel()} of the interface
\code{Parcable} to define how they can be encoded using \code{Parcel}'s
standard marshalling methods. We can invoke \code{createFromParcel()}
via reflection in most cases, and are thus able to cover most argument
types appearing in method calls. This is crucial, as once we miss
unmarshalling one argument for a method call, we cannot access remaining
arguments in the rest of \code{Parcel} object's byte array.

\paragraph{Kernel/User-Space Communication}


As event interception takes place solely inside the kernel space and
unmarshalling relies on Android's Java API, we need a mechanism that
allows us to pass data from inside the kernel space up to an app.
Moreover, we need some means to let the user control the kernel module
for even the most basic tasks, for example, to switch event interception
on or off from an app.
However, Android has no built-in way to serve as a solution, but we were
able to use
\code{netlink},\footnote{http://www.linuxfoundation.org/collaborate/workgroups/\\networking/generic\_netlink\_howto}
which is a socket based mechanism of the Linux kernel that can
bidirectionally communicate with user space. We placed the
implementation of the communication endpoints into our kernel module and
into our app, and thus, leave the Android framework as such completely
unmodified.
As the Android API does not offer \code{netlink} support, we had to
build a custom \code{netlink} endpoint for our app, using the
\emph{Netlink Protocol Library Suite}
(\code{libnl}\footnote{http://www.carisma.slowglass.com/~tgr/libnl/}).
It runs as a shared C++-library, \code{droidtracer.so} (see
Fig.~\ref{fig:rmi}), in the same process as our app. We had to extract
the core functionality for \code{netlink} from \code{libnl} and
recompile it for Android devices using Android's Native Developer Kit
(NDK).
\code{Netlink} allows to declare a callback method to receive kernel
data, so that \code{droidtracer.so} does not have to poll our kernel
module for events. As the ultimate goal is to transmit them all the way
to \code{DroidTracerService} we use reflection and Java Native Interface
(JNI) to register a method in Java that is automatically triggered if
we forward data received from \code{netlink} to it.

%





\section{Use Cases---Runtime Verification}
\label{sec:rv}
As already pointed out in \S~\ref{sec:intro}, there is a myriad of
reasons why users may want to closely monitor their mobile devices.
Although our add-on is application-agnostic, our motivation is to bring
techniques developed in the area of runtime verification (cf.\
\cite{Chen:2009:PTS:1532891.1532921,BKV2013}) closer to Android.  In a
nutshell, runtime verification subsumes tools and techniques that aim at
checking that a system's actual behaviour agrees to a (formal)
specification of pre-defined behaviour.  For example, one could specify
that ``no installed app should ever send an SMS to a phone number, which
is not stored in the user's contacts list,'' using a dedicated temporal
logic, for example.
Arguably, a major focus in runtime verification research right now is to
be able to monitor not just reactive, but also data-intensive systems,
where one not only needs to know that a message was sent, but also by
who and to whom (cf.\ keynote speech ``Runtime Verification with Data''
by M.\ Leucker at the Runtime Verification conference 2013, Rennes,
France).  Temporal logics and regular expressions, traditionally used in
that domain are typically not sufficient to express such parameterised
behavioural policies, and if they are (cf.\
\cite{Chen:2009:PTS:1532891.1532921}), there is usually no tool support
for the domain that concerns us.

A recent paper \cite{BKV2013}, introduces a general purpose monitoring
algorithm for a parameterised logic, which we have implemented as a
client that directly utilises the API of the DroidTracer add-on.
The above policy can be formalised in this logic and be applied to any
app installed on the phone.  To that end, our client provides a
graphical front-end that lists all the installed apps of a device and
users can choose which ones are being monitored simply by clicking on
its name in a list.

Although our approach is platform-centric, we give users the possibility
to exempt certain apps from being monitored, mainly for two reasons:
firstly, to avoid false negatives in case certain apps are more
trustworthy than others, and secondly, for performance reasons, since
our runtime verification client, internally, instantiates one runtime
verification monitor per app and policy.\footnote{In the worst-case,
  monitors are of exponential size wrt.\ a policy, depending on the
  policy as well as concretely observed system behaviour
  \cite{BKV2013}.}  Therefore, the fewer apps are being monitored, the
less overhead and disruption for the rest of the system.


Note that, in an abstract sense, the aforementioned tools such as
AppGuard are also runtime verification tools.  However, rather than
employing tools and techniques developed in the area of runtime
verification directly, they use alternative means to define security
policies (in case of AppGuard, for example, \emph{security automata}
that have been introduced by Schneider
\cite{DBLP:journals/tissec/Schneider00}).  Also, off the shelf security
automata are not suitable as a fully declarative policy specification
language over potentially infinite data domains (e.g., set of all IPs,
email addresses, etc.).  



\section{Performance Overhead}
\label{sec:exps}
\begin{table*}[ht]
  \begin{center} {\footnotesize
      \caption{
        Execution of Android API method calls (each up to 10,000 times) with and without DroidTracer. The margin
        of error is given for the 95\% confidence interval.}
    \label{tbl:exps}
\scalebox{1.01}{%
\begin{tabular}{llrrrrr}
  \toprule
  \multicolumn{1}{l}{Interface} & \multicolumn{1}{l}{Method} & \multicolumn{1}{r}{Android} & \multicolumn{1}{r}{Kprobes} &
  \multicolumn{1}{r}{Add-on} & \multicolumn{1}{r}{Kprobes}  & \multicolumn{1}{r}{Add-on}\\
  \multicolumn{1}{c}{} & \multicolumn{1}{c}{} & \multicolumn{1}{r}{(in ms)} & \multicolumn{1}{r}{(in ms)} &
  \multicolumn{1}{r}{(in ms)} & \multicolumn{1}{r}{(Overhead in \%)}  & \multicolumn{1}{r}{(Overhead in \%)}\\\midrule
  \rowcol TelephonyManager & getDeviceId & 5309 $\pm$ 15 & 5517 $\pm$ 18 & 5811 $\pm$ 11 & 3.92 & 9.46\\[0ex]
  TelephonyManager & getSimSerialNumber & 5346 $\pm$ 16 & 5524 $\pm$ 16 & 5817 $\pm$ 7 &  & 8.81 \\[0ex]
  \rowcol LocationManager & getLastKnownLocation & 3516 $\pm$ 13 & 3562 $\pm$ 13 & 4126 $\pm$ 5 &  & 17.35\\[0ex]
  SmsManager & sendTextMessage & 9166 $\pm$ 13 & 9396 $\pm$ 13 & 10216 $\pm$ 10 & 2.51 & 11.46\\[0ex]
  \rowcol PackageManager & getInstalledApplications & 15730 $\pm$ 204 & 15514 $\pm$ 202 & 15422 $\pm$ 172 &  &  \\[0ex] 
  ConnectivityManager & getAllNetworkInfo & 5769 $\pm$ 53 & 5841 $\pm$ 60 & 5671 $\pm$ 7 &  &  \\[0ex]
  \rowcol BufferedReader & readLine & 15360 $\pm$ 72 & 15531 $\pm$ 67 & 15455 $\pm$ 38 &  &  \\[0ex]
 \bottomrule
\end{tabular} }}
\end{center}
\vspace{-1em}
\label{turns}
\end{table*}
When our add-on is being executed with our runtime verification client
on, the average performance overhead is 38.6\%. This was
  determined on a Nexus 7 (1st generation) tablet with a quad-core CPU
  and 1 GB RAM running Android OS 4.3. We used seven test apps,
  specifically written for our purpose, i.e., each of the apps was
  designed to generate 100 runs of up to 10,000 events of the following
  sorts: 1.\ access the device ID (IMEI), 2.\ read the SIM card serial
  number, 3.\ request location, 4.\ send SMS, 5.\ look up list of
  installed apps, 6.\ access connection status about all network types,
  and 7.\ read file from SD card.  Besides monitoring the policy
introduced in \S\ref{sec:rv}, namely that no app shall send a text
message to a number not recorded in the user's contacts list, we have
instantiated the policies used in \cite{BKV2013} to cover a somewhat
broader range of possible applications.  However, as the results of
these test runs are specific only to our implementation of runtime
verification, we also need to measure the performance overhead of our
``bare bones'' add-on when no further analysis is undertaken.
Table~\ref{tbl:exps}, shows the execution time when intercepting the API
method calls of the above eight events in three different modes of
operation: First, we ran the test apps without our add-on enabled to get
a reference execution time for the unmodified system. Second, we enabled
only the event interception part of our add-on but no further event
processing, and thirdly, we added unmarshalling and netlink
communication to it in order to determine whether or not there is a
performance penalty associated with how we process the intercepted
information back in user space.


%

As the results show, the actual performance overhead of using just our
kernel module with \code{kprobes} is only 2.51--3.92\%, whereas
the complete performance overhead, when including netlink and
unmarshalling, is 8.81--17.35\%. During the experiments, we
  ran all four cores of the Nexus 7 on a fixed frequency rate using the
  performance governor, which allowed us to reduce the margin of error
  dramatically. Note that we left cells empty in Table~\ref{tbl:exps},
  where overhead could not significantly be determined wrt. the t-test
  for the 95\% confidence interval.
What is noteworthy is that \code{getDeviceId()} and
\code{getSimSerialNumber()} have significant lower overhead than
\code{getLastKnownLocation()} and \code{sendTextMessage()}, as both
former method signatures have no arguments that require unmarshalling. 
%
%
The call \code{getLastKnownLocation()} has the highest overhead. A
probable explanation seems to be that its arguments contain several
complex objects, for example, one of type \code{LocationRequest}. Their
unmarshalling involves additional reflection calls; that is, to find and
invoke the method \code{createFromParcel()}, which a class has to
implement, if it wants to be sent through Binder. As
  \code{sendTextMessage()} contains only Java primitives as arguments,
  its unmarshalling overhead is slightly lower.

\section{Conclusions and Future Work}
\label{sec:conc}
Our benchmarks suggest that we sit comfortably between the competition.
For example, Aurasium's interception of \code{getDeviceId()} has a 35\%
and \code{getLastLocation()} a 34\% performance overhead \cite{XSA2012},
whereas AppGuard's benchmarks range between 0.8 and 21.4\% overhead.
TaintDroid, although more of a taint flow analysis rather than a pure
monitoring tool, has a performance overhead of up to 29\%, which is in
the range of our measurements of 38.6\% average overhead with the
runtime verification client enabled.
However, one should keep in mind that unlike the other solutions,
DroidTracer does not rely on app or system modifications and therefore
has limited potential for optimisation.  The fact that, in particular,
our kernel module performs well under these constraints, is somewhat
surprising but largely due to our ability to intercept all the relevant
data at a single point of entry, i.e., the Binder.
More so, we have shown that detailed systems monitoring on Android is
possible even without any modifications to the platform itself.

However, this is preliminary work, a proof of concept.  While we are
content with the event interception part of our work and already offer
an API, such that others can use and benefit from it as is, we have yet
to substantiate our claim that runtime verification as outlined in
particular in \S~\ref{sec:rv} really is beneficial to system security.
Works that analyse current threats and security trends in that area
(cf.\ \cite{ZJ2012, FFC+2011, EOMC2011}) seem to suggest that the total
number of Android attacks is on the rise, but that most of the threats
follow only a handful of different patterns.  Our working hypothesis is
that these patterns can and should be translated into (temporal logic)
policies, which then in turn are monitored by our tool.  The essential
ingredients for this undertaking were presented in this paper and shown
to work.


\section{Acknowledgments}

NICTA is funded by the Australian Government as represented by the
Department of Broadband, Communications and the Digital Economy and the
Australian Research Council through the ICT Centre of Excellence
program.

{\footnotesize \bibliographystyle{acm}
\bibliography{bibliography}}

\begin{thebibliography}{10}

\bibitem{BGH+2013}
{\sc Backes, M., Gerling, S., Hammer, C., Maffei, M., and von Styp-Rekowsky,
  P.}
\newblock App{G}uard - {F}ine-grained policy enforcement for untrusted
  {A}ndroid applications.
\newblock Tech. Rep. A/02/2013, Saarland University, 2013.

\bibitem{BKV2013}
{\sc Bauer, A., K{\"u}ster, J.-C., and Vegliach, G.}
\newblock From propositional to first-order monitoring.
\newblock In {\em RV\/} (2013), vol.~8174 of {\em LNCS}, Springer.

\bibitem{Bugiel2012}
{\sc Bugiel, S., Davi, L., Dmitrienko, A., Fischer, T., Sadeghi, A.-R., and
  Shastry, B.}
\newblock Towards taming privilege-escalation attacks on {A}ndroid.
\newblock In {\em NDSS\/} (2012), The Internet Society.

\bibitem{Chen:2009:PTS:1532891.1532921}
{\sc Chen, F., and Ro\c{s}u, G.}
\newblock Parametric trace slicing and monitoring.
\newblock In {\em TACAS\/} (2009), vol.~5505 of {\em LNCS}, Springer.

\bibitem{EGC+2010}
{\sc Enck, W., Gilbert, P., Chun, B.-G., Cox, L.~P., Jung, J., McDaniel, P.,
  and Sheth, A.~N.}
\newblock Taintdroid: an information-flow tracking system for realtime privacy
  monitoring on smartphones.
\newblock In {\em OSDI\/} (2010), USENIX.

\bibitem{EOMC2011}
{\sc Enck, W., Octeau, D., McDaniel, P., and Chaudhuri, S.}
\newblock A study of {A}ndroid application security.
\newblock In {\em USENIX Security Symposium\/} (2011), USENIX.

\bibitem{FFC+2011}
{\sc Felt, A.~P., Finifter, M., Chin, E., Hanna, S., and Wagner, D.}
\newblock A survey of mobile malware in the wild.
\newblock In {\em SPSM\/} (2011), ACM.

\bibitem{DBLP:journals/corr/GunadiT13}
{\sc Gunadi, H., and Tiu, A.}
\newblock Efficient runtime monitoring with metric temporal logic: A case study
  in the {A}ndroid operating system.
\newblock {\em CoRR abs/1311.2362\/} (2013).

\bibitem{HZF+2012}
{\sc Han, D., Zhang, C., Fan, X., Hindle, A., Wong, K., and Stroulia, E.}
\newblock Understanding {A}ndroid fragmentation with topic analysis of
  vendor-specific bugs.
\newblock In {\em WCRE\/} (2012), IEEE.

\bibitem{Hornyack2011}
{\sc Hornyack, P., Han, S., Jung, J., Schechter, S., and Wetherall, D.}
\newblock These aren't the droids you're looking for: retrofitting {A}ndroid to
  protect data from imperious applications.
\newblock In {\em CCS\/} (2011), ACM.

\bibitem{NKZ2010}
{\sc Nauman, M., Khan, S., and Zhang, X.}
\newblock Apex: extending android permission model and enforcement with
  user-defined runtime constraints.
\newblock In {\em ASIACCS\/} (2010), ACM.

\bibitem{DBLP:journals/tissec/Schneider00}
{\sc Schneider, F.~B.}
\newblock Enforceable security policies.
\newblock {\em ACM Trans. Inf. Syst. Secur. 3}, 1 (2000).

\bibitem{XSA2012}
{\sc Xu, R., Sa\"{\i}di, H., and Anderson, R.}
\newblock Aurasium: practical policy enforcement for {A}ndroid applications.
\newblock In {\em USENIX Security Symposium\/} (2012), USENIX.

\bibitem{Yoon2012}
{\sc Yoon, C., Kim, D., Jung, W., Kang, C., and Cha, H.}
\newblock {AppScope}: Application energy metering framework for {A}ndroid
  smartphone using kernel activity monitoring.
\newblock In {\em USENIX ATC\/} (2012), USENIX.

\bibitem{ZJ2012}
{\sc Zhou, Y., and Jiang, X.}
\newblock Dissecting {A}ndroid malware: Characterization and evolution.
\newblock In {\em SSP\/} (2012), IEEE.

\end{thebibliography}


\end{document}